\documentstyle[editedvolume,graphicx,numreferences,epsfig]{crckapb} 

\def\Journal#1#2#3#4{(#4){#1} {\bf #2}, pp.#3}

\def\NCA{\em Nuovo Cimento}

\def\PLB{{\em Phys. Lett.}  B}

\def\PRD{{\em Phys. Rev.} D}
\def\ZPC{{\em Z. Phys.} C}
\def\EPG{{\em Eur.Phys.J.} C}

\def\AOP{{\em Ann. of Phys. (NY)}}

\def\be{\begin{equation}}
\def\ee{\end{equation}}
\def\bea{\begin{eqnarray}}
\def\eea{\end{eqnarray}}
\def \Pom {{\hspace{ -0.05em}I\hspace{-0.25em}P}}
\def \Reg {{\hspace{ -0.05em}I\hspace{-0.25em}R}}

\begin{opening}
\title{
VECTOR MESON PHOTOPRODUCTION IN
THE SOFT DIPOLE POMERON MODEL FRAMEWORK
}

\author{\underline{A. PROKUDIN $^{1,2,a}$}}
\author{E. MARTYNOV$^{3,4,b}$}
\author{E. PREDAZZI$^{1,c}$}
\institute{
$^1$  Dipartimento di Fisica Teorica,
Universit\`a Degli Studi Di Torino, 
Via Pietro Giuria 1,
10125 Torino, 
ITALY
and
Sezione INFN di Torino,
 ITALY\\
$^2$ Institute For High Energy Physics,
142281 Protvino, RUSSIA\\
$^3$ Bogolyubov Institute for Theoretical Physics, National \\ Academy of
Sciences of Ukraine, \\ 03143 Kiev-143, Metrologicheskaja 14b, UKRAINE \\
$^4$  Institut de physique Bat B5-a 
Universit\'e de Li\`ege 
Sart Tilman B-4000 Li\`ege, 
  BELGIQUE\\
{\small {\rm (A)} E-mail: prokudin@to.infn.it 
{\rm (B)} E-mail:
e.martynov@guest.ulg.ac.be 
{\rm (C)} E-mail: predazzi@to.infn.it}
}

\end{opening}

\runningtitle{VECTOR MESON PHOTOPRODUCTION...}

\begin{document}

\begin{abstract}
Exclusive photoproduction of all vector mesons by real and
virtual photons is considered in the Soft Dipole Pomeron model.
It is emphasized that being the Pomeron in this model a double
Regge pole with intercept equal to one, we are led to rising
cross-sections but the unitarity bounds are not violated. It is
shown that all available data for $\rho, \omega, \varphi, J/\psi
$ and $\Upsilon $ in the region of energies 1.7 $\leq W \leq $
250 GeV and photon virtualities 0 $\leq Q^2 \leq $ 35 GeV$^2$ ,
including the differential cross-sections in the region of
transfer momenta
 0 $\leq |t| \leq$ 1.6 GeV$^2$, are well described
by the model.
\end{abstract}
\vspace*{-1cm}


\section{Introduction}
A new precise measurement of $J/\psi$ exclusive photoproduction
by ZEUS \cite{NEWZEUS} opens a new window in our understanding
 of the process and allows us to give more
accurate predictions for future experiments.

The key issue of the dataset \cite{NEWZEUS} is the diffractive
cone shrinkage observed in  $J/\psi$ photoproduction which
leads us to consider it a soft rather than pure QCD
process so that we can apply the Soft
Dipole Pomeron exchange \cite{owrmodel} model.

\begin{figure}[h]
\begin{center}
\includegraphics[scale=0.35]{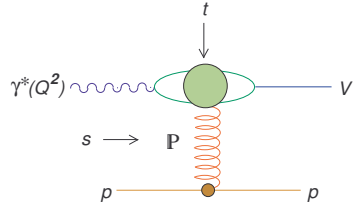}
\caption{Photoproduction of a vector meson.}
\label{Figure 1}
\end{center}
\end{figure}

The basic diagram is depicted in Figure \ref{Figure 1}; $s$ and $t$ are
the usual Mandelstam variables, $Q^2=-q^2$ is the virtuality of the
photon.

It is well known that high-energy representation of the
scattering amplitude may be expressed as a sum over the
the appropriate Regge poles in the complex $j$ plane
\cite{ref:PredazziBarone}

\be
A(s,t)_{s\to \infty}\approx \sum\limits_{i}
\eta_{i}(t)\beta_{i}(t)(\cos\theta_{t})^{\alpha_{i}(t)},
\ee
where $\eta_{i}(t)$ is the signature factor and $\theta_t$ is the scattering angle
in the $t$ channel.

In the case of vector meson photoproduction we utilize the
variable $z \sim \cos \theta_t$ \be z =
\frac{2(W^2-M_p^2)+t+Q^2-M_V^2}{\sqrt{(t+Q^2-M_V^2)^2+4M_V^2Q^2}}
\ee where  $W^2=(p+q)^2\equiv s$, $M_V$ is the vector meson
mass, $M_p$ is the proton mass.

 Assuming vector meson dominance \cite{ref:vdm}, the relation between
the forward cross sections of $\gamma p\rightarrow V p$ and $V
p\rightarrow V p$ is given by \be
\frac{d\sigma}{dt}(t=0)_{\gamma p\rightarrow V p}=
\frac{4\pi\alpha}{f_V^2}\frac{d\sigma}{dt}(t=0)_{V p\rightarrow
V p} \ee where the strength of the vector meson coupling
$\frac{4\pi}{f_V^2}$ may be found from $e^+e^-$ decay width of
vector meson $V$ \be \Gamma_{V\rightarrow e^+e^-}=
\frac{\alpha^2}{3}\frac{4\pi}{f_V^2} m_V \ee 

Using the property $\Gamma_{V\rightarrow e^+e^-}/<Q^2_j>\simeq const$
we can obtain the following
approximate relations

\be
{m_\rho/f_\rho^2}:{m_{\omega}/f_{\omega}^2}:{m_{\varphi}/f_{\varphi}^2}:
{m_{J/\psi}/f_{J/\psi}^2} =
{9}:{1}:{2}:{8}\;
\ee
which are in fairly good agreement with experimental measurements
of decay widths \cite{ref:groom}.

We take into account these relations by introducing coefficients
$N_V$ (following to \cite{ref:Nemchik} ) and writing the
amplitude as $A_{\gamma p\rightarrow V p}=N_C N_V A_{V
p\rightarrow V p}$, where \be N_C = 3\: ; N_\rho =
\frac{1}{\sqrt{2}}\: ; N_\omega = \frac{1}{3\sqrt{2}}\: ; N_\phi
= \frac{1}{3}\: ; N_{J/\psi} = \frac{2}{3}\: . \ee

The
amplitude of the process $V p\rightarrow V p$
may be written in the
following form \be
A(z,t;M_V^2,\tilde Q^2) = \Pom (z,t;M_V^2, \tilde Q^2) + f(z,t;
M_V^2,\tilde Q^2) + ...\;, \ee where,
$\tilde Q^{2} = Q^{2}+M_V^2$.  

$\Pom (z,t;M_V^2, \tilde
Q^2)$ is the Pomeron contribution for which we use the so called
dipole Pomeron which gives a very good description of all
hadron-hadron total cross
sections \cite{ref:dipole_hadron},\cite{ref:COMPETE}.
Specifically, $\Pom$ is given by  \cite{ref:JMP} 
\bea 
\Pom
(z,t;M_V^2, \tilde Q^2) = ig_{0}(t;M_V^2, \tilde Q^2)
(-iz)^{\alpha_{\Pom}(t)-1} +  \\ \nonumber
ig_{1}(t;M_V^2, \tilde Q^2)ln(-iz)
(-iz)^{\alpha_{\Pom}(t)-1}\; , 
\label{eq:pomeron} 
\eea where the
first term is a single $j$-pole contribution and the second
(with an additional $ln(-iz)$ factor) is the contribution of the
double $j$-pole.

A similar
expression applies to the contribution of the $f$-Reggeon 
\be
f(z,t;M_V^2, \tilde Q^2) = ig_{f}(t;M_V^2,\tilde Q^2)
(-iz)^{\alpha_{f}(t)-1}. 
\label{eq:reggeon} 
\ee

It is important to stress that in this model the intercept of
the Pomeron trajectory is equal to 1 \be \alpha_{\Pom}(0) = 1.
\ee Thus the model does not violate the Froissart-Martin bound
~\cite{ref:MartinF}.
 
For $\rho$ and $\varphi$ meson photoproduction we write the
scattering amplitude as the sum of a Pomeron and $f$
contribution. According to the Okubo-Zweig rule, the $f$ meson
contribution should be suppressed in the production of
the $\varphi$ and $J/\psi$ mesons, but given the
present crudeness of the state of the art, we added the $f$
meson contribution in the $\varphi$ meson case.

For $\omega$ meson photoproduction, we include also $\pi$ meson exchange
(see also the discussion in \cite{ref:DL}), which is needed in the
 low energy sector given that we try to describe the data for all
energies $W$. Even though we did not expect it,
the model 
describes well the data down to threshold.

In the integrated elastic cross section \be \displaystyle
\sigma(z, M_V^2, \tilde Q^2)^{\gamma p\rightarrow Vp}_{el} =
4\pi\int\limits_{t_{-}}^{t_{+}}dt|A^{\gamma p\rightarrow
Vp}(z,t;M_V^2,\tilde Q^2)|^{2}\; , \label{eq:sigma} \ee
 the upper and lower limits
 \be
2t_{\pm}=\pm
\frac{L_{1}L_{2}}{W^{2}}-(W^{2}+Q^{2}-M_{V}^{2}-2M_{p}^{2})+
\frac{(Q^{2}+M_{p}^{2})(M_{V}^{2}-M_{p}^{2})}{W^{2}}, \ee \be
L_{1}=\lambda(W^{2},-Q^{2},M_{p}^{2}),\qquad
L_{2}=\lambda(W^{2},M_{V}^{2},M_{p}^{2}), \ee \be
\lambda^{2}(x,y,z)=x^{2}+y^{2}+z^{2}-2xy-2yz-2zx, \ee are
determined by the kinematical condition $-1\leq
\cos\theta_{s}\leq 1$ where $\theta_{s}$ is the scattering angle
in the s-channel of the process.

For the Pomeron contribution (\ref{eq:pomeron}) we use a
nonlinear trajectory \be\label{eq:trajectory_of_pomeron}
\alpha_\Pom (t)=1+\gamma (\sqrt{4m_\pi^2}-\sqrt{4m_\pi^2-t}\:),
\ee where $m_\pi$ is the pion mass. Such a trajectory was
utilized for photoproduction amplitudes in
\cite{ref:dipole_vector}, \cite{ref:Jenk} and its roots are very old
\cite{ref:Enrico1965}.

For the $f$-meson contribution for the sake of simplicity we use
the standard linear Reggeon trajectory \be \alpha_\Reg (t)=\alpha_\Reg
(0)+\alpha'_\Reg (0)\: t \: . \ee 

In the case of nonzero virtuality of the photon, we have a new
variable in play $Q^2=-q^2$. At the same time, the cross section 
$\sigma_L$ is nonzero.

\section{The Model}
For the Pomeron residues we use the following parametrization
\bea\label{eq:couplings} g_i(t; M_V^2,\tilde
Q^2)=\frac{g_i}{Q_i^2+\tilde Q^2}
exp(b_i(t; \tilde Q^2))\; , \\
\nonumber
i=0,1 \: .
\nonumber
\eea
\noindent The slopes are chosen as
\bea\label{eq:slopes}
b_i(t; \tilde Q^2)=\Bigl(b_{i0}+\frac{b_{i1}}{1+\tilde Q^2/Q_{b}^2}\Bigr)(\sqrt{4m_\pi^2}-\sqrt{4m_\pi^2-t}\:)\; , \\
\nonumber i=0,1 \: , \eea to comply with the previous choice
(\ref{eq:trajectory_of_pomeron}) and analyticity requirements
\cite{ref:Enrico1965}.

\noindent The Reggeon residue is \be\label{eq:couplingsR}
g_\Reg(t; M_V^2, \tilde Q^2)=\frac{g_\Reg
M_p^2}{(Q_\Reg^2+\tilde Q^2)\tilde Q^2}exp(b_\Reg(t; \tilde
Q^2))\; , \ee where \be\label{eq:slopesR} b_\Reg(t; \tilde
Q^2)=\frac{b_\Reg}{1+\tilde Q^2/Q_{b}^2}t\; , \ee $g_0, \; g_1$,
$Q_0^2\; (GeV^2)$, $ Q_1^2\; (GeV^2)$, $ Q_\Reg^2\; (GeV^2)$, $
Q_b^2\; (GeV^2)$, $b_{00}\; (GeV^{-1})$, $b_{01}\; (GeV^{-1})$,
$b_{10}\; (GeV^{-1})$, $b_{11}\; (GeV^{-1})$, $b_{\Reg}\;
(GeV^{-2})$ are adjustable parameters. $\Reg=f$ for $\rho$ and
$\varphi$, $\Reg=f,\pi$ for $\omega$. We use the same slope
$b_\Reg$ for $f$ and $\pi$ Reggeon exchanges.

\subsection{Photoproduction of vector mesons by real photons ($Q^2=0$).}

In the fit we use all available data starting from the threshold
for each meson. As the new dataset of ZEUS \cite{NEWZEUS} allows us to explore
the effects of nonlinearity of the Pomeron trajectory and residues.
In the region of non zero $Q^2$ the combined
data of H1 and ZEUS is used.

The whole  set of data is composed of $357$  experimental points
\footnote{The data are available at \\
REACTION DATA Database {\it http://durpdg.dur.ac.uk/hepdata/reac.html} \\
CROSS SECTIONS PPDS database {\it http://wwwppds.ihep.su:8001/c1-5A.html}}
 and, with
a grand total of $12$ parameters, we find $\chi^2/{\rm d.o.f}=1.49$. The
main contribution to $\chi^2$ comes from the low energy region ( $W\le
4\;GeV$); had we started fitting from $W_{min}=4\;GeV$, the resulting
$\chi^2/{\rm d.o.f}=0.85$ for the elastic cross sections  would be much better and more appropriate for a
high energy model.

In order to get a reliable description and the parameters
of the trajectories and residues we use elastic cross sections
for each process from threshold  up to the highest
values of the energy and  differential cross sections in the whole t-region 
where data are available: 0 $\le |t| \le$ 1.6 $GeV^2$.

The parameters are given in \cite{ref:OURPAPER}.

\begin{figure}[h]
\begin{center} 
{\vspace*{ -1cm} \epsfxsize=70mm
\epsffile{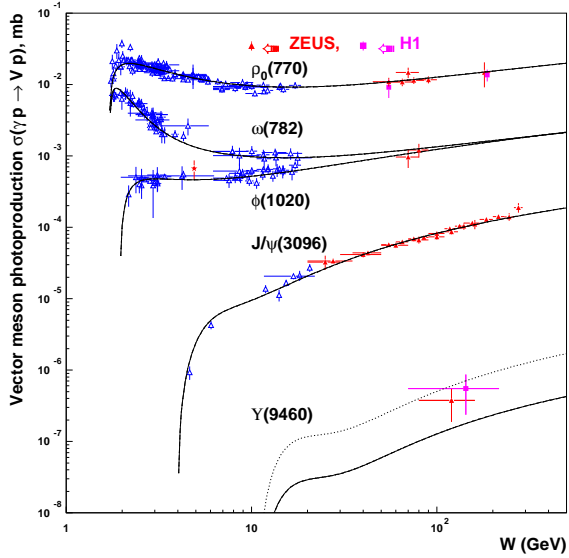}} 
\vskip-1.cm \caption{Elastic
cross-sections for vector meson photoproduction. The solid curve for
$\Upsilon (9460)$ production corresponds to
$N_\Upsilon=N_\varphi$, the dotted line to $N_\Upsilon=N_{J/\psi}$.
\label{fig:mesons}}
\end{center}
\end{figure}

The results are presented in Fig. \ref{fig:mesons}, which shows also the
 prediction of the model for $\Upsilon (9460)$ photoproduction.

\subsection{Photoproduction of vector mesons by virtual photons\\ ($Q^2>0$).}

In (\ref{eq:couplings}) and (\ref{eq:couplingsR}) the $Q^2$-
dependence ($\tilde Q^2=Q^2+M_V^2$) is completely fixed up to an
{\it a priori} arbitrary dimensionless function $f(Q^2)$ such
that $f(0)=1$. Thus, we may introduce a new factor that
differentiates virtual from real photoproduction:

\be\label{eq:fpom}
f(Q^2)=\Big(\frac{M_V^2}{\tilde Q^2}\Big)^n
\ee

Accordingly, in the case $Q^2\neq 0$ we use the following parametrizations
for Pomeron couplings (compare with Eq. \ref{eq:couplings}):
\be\label{eq:couplingsqP}
\hat g_i(t; \tilde Q^2, M_V^2)= f(Q^2)g_i(t; \tilde Q^2, M_V^2),\; i=0,1,
\ee
where, for the sake of completeness, we will examine three different
{\it choices} for the asymptotic $Q^2$ behaviour of the Pomeron
residue

\noindent
{\it \underline{Choice I}}
 \be\label{eq:couplingsChoice}
n=1, \qquad
 \sigma_T({Q^2 \rightarrow \infty}) \sim \frac{1}{Q^8}\; .
\ee {\it \underline{Choice II}}
 \be \label{eq:couplings1}
n=0.5, \qquad
 \sigma_T({Q^2 \rightarrow \infty}) \sim \frac{1}{Q^6}\; .
\ee
{\it \underline{Choice III}}
 \be \label{eq:couplings2}
n=0.25, \qquad
 \sigma_T({Q^2 \rightarrow \infty}) \sim \frac{1}{Q^5}\; .
\ee

\noindent For the reggeon couplings we have
\be
f_\Reg(Q^2)=\Big(\frac{c_1 M_V^2}{c_1 M_V^2+Q^2}\Big)^{n_2}\; ,
\ee
where $c_1$ is an adjustable parameter and
$n_2=0.25, \; -0.25, \; -0.5$ for {\it choice I, II, III}.

Accordingly, in the case $Q^2\neq 0$ we use the following parametrizations
for Reggeons couplings (compare with Eq. \ref{eq:couplingsR}):
\be\label{eq:couplingsq}
\hat g_\Reg(t; \tilde Q^2, M_V^2)= f_\Reg(Q^2)g_\Reg(t; \tilde Q^2, M_V^2)\; .
\ee

The lack of data on the ratio $\sigma_L/\sigma_T$, especially in
the high $Q^2$ domain, does not allow us to draw definite
conclusions about its asymptotic behaviour. There may 
be several realizations of the model with
different asymptotic behaviour of $\sigma_L/\sigma_T$
\cite{owrmodel}. As a demonstration of such a possibility we
use the following
(most economical) parametrization for $R$ (which cannot be
deduced from the Regge theory)

\noindent{\it \underline{Choice I, II, III}}
\be R(Q^2, M_V^2) =
\Big(\frac{c M_V^2+Q^2}{c M_V^2}\Big)^{n_1}-1
\label{eq:ratio}
\ee
where $c$ and $n_1$
are adjustable parameters for {\it choice I, II, III}.

We have, thus, 3 additional adjustable parameters as compared with
real photoproduction. In order to obtain the values of
the parameters for the case $Q^2\ne 0$,
we fit just the data\footnote{The data are available at \\
REACTION DATA Database {\it http://durpdg.dur.ac.uk/hepdata/reac.html} \\
CROSS SECTIONS PPDS database {\it http://wwwppds.ihep.su:8001/c1-5A.html}}
on $\rho_0$ meson photoproduction in the region $0\le Q^2\le 35 \; GeV^2$;
the parameters for photoproduction by real photons are fixed. 

The parameters thus obtained may be found in \cite{ref:OURPAPER}.

\begin{figure}[h]
\centering {\vspace*{ -1cm} \epsfxsize=70mm
\epsffile{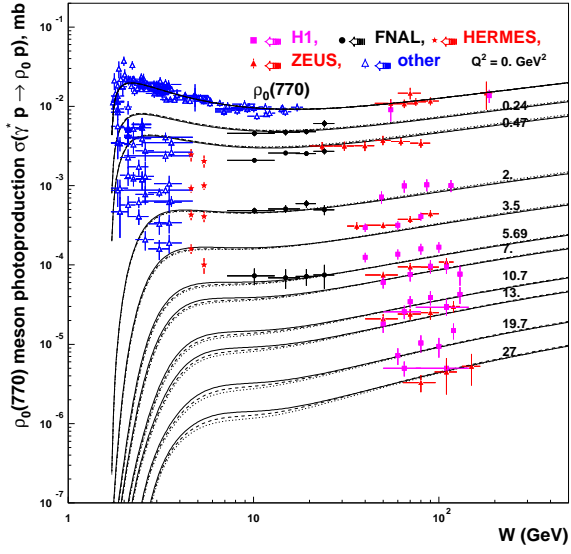}} 
\vskip-1.cm \caption{Elastic cross section of exclusive
$\rho_0$ virtual photoproduction as a function
of $W$ for different values of $Q^2$.
\label{fig:rho}}
\end{figure}

The results of the fit are depicted in Fig. \ref{fig:rho}. In this figure as well as in all following ones the solid lines,
dashed
lines and dotted lines correspond to the {\it choice I, II, III}
correspondingly.

\begin{figure}[h]
\parbox[c]{5.cm}{\epsfxsize=50mm
\epsffile{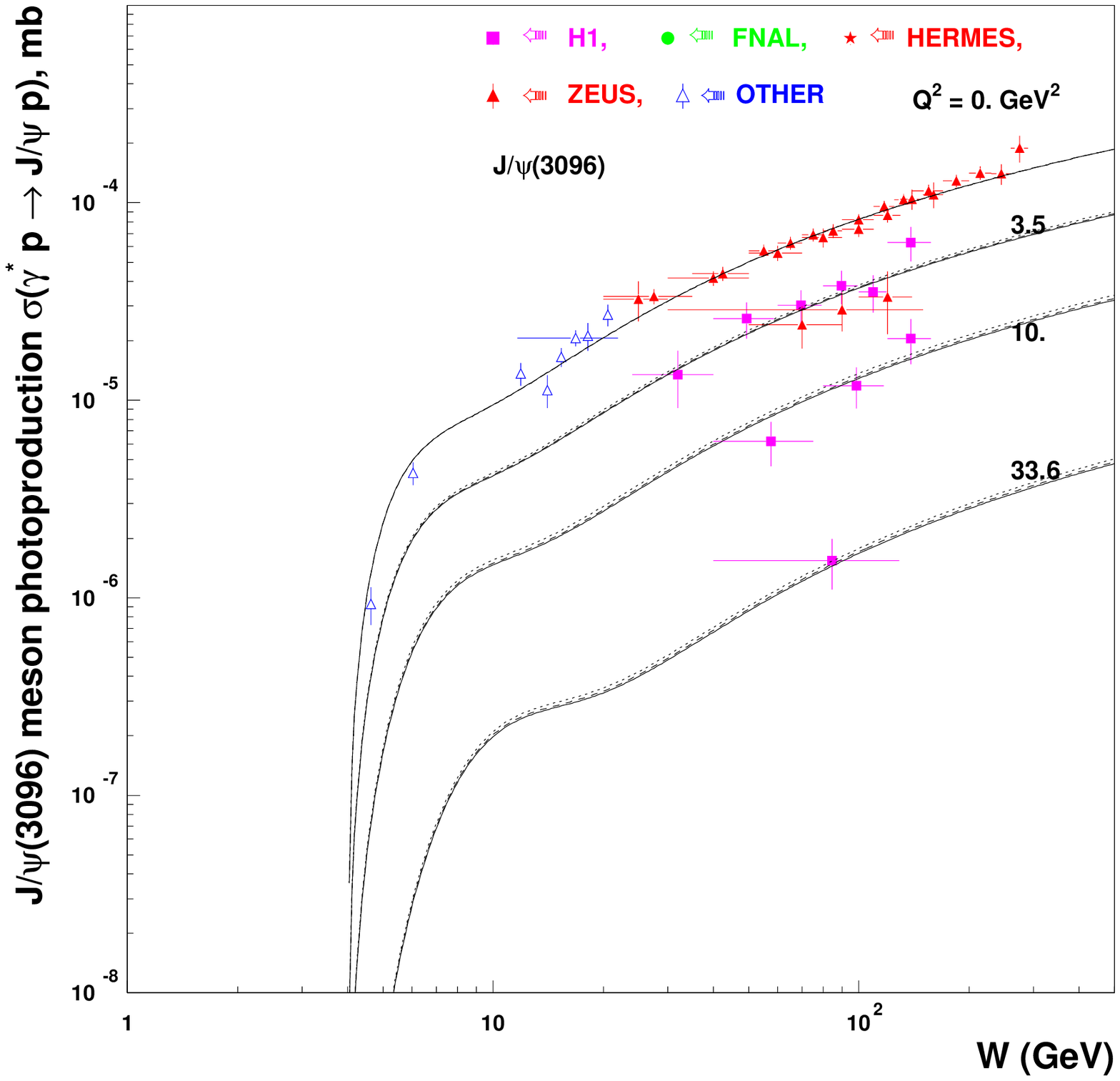}} \hfill~\parbox[c]{5.cm}{\epsfxsize=50mm
\epsffile{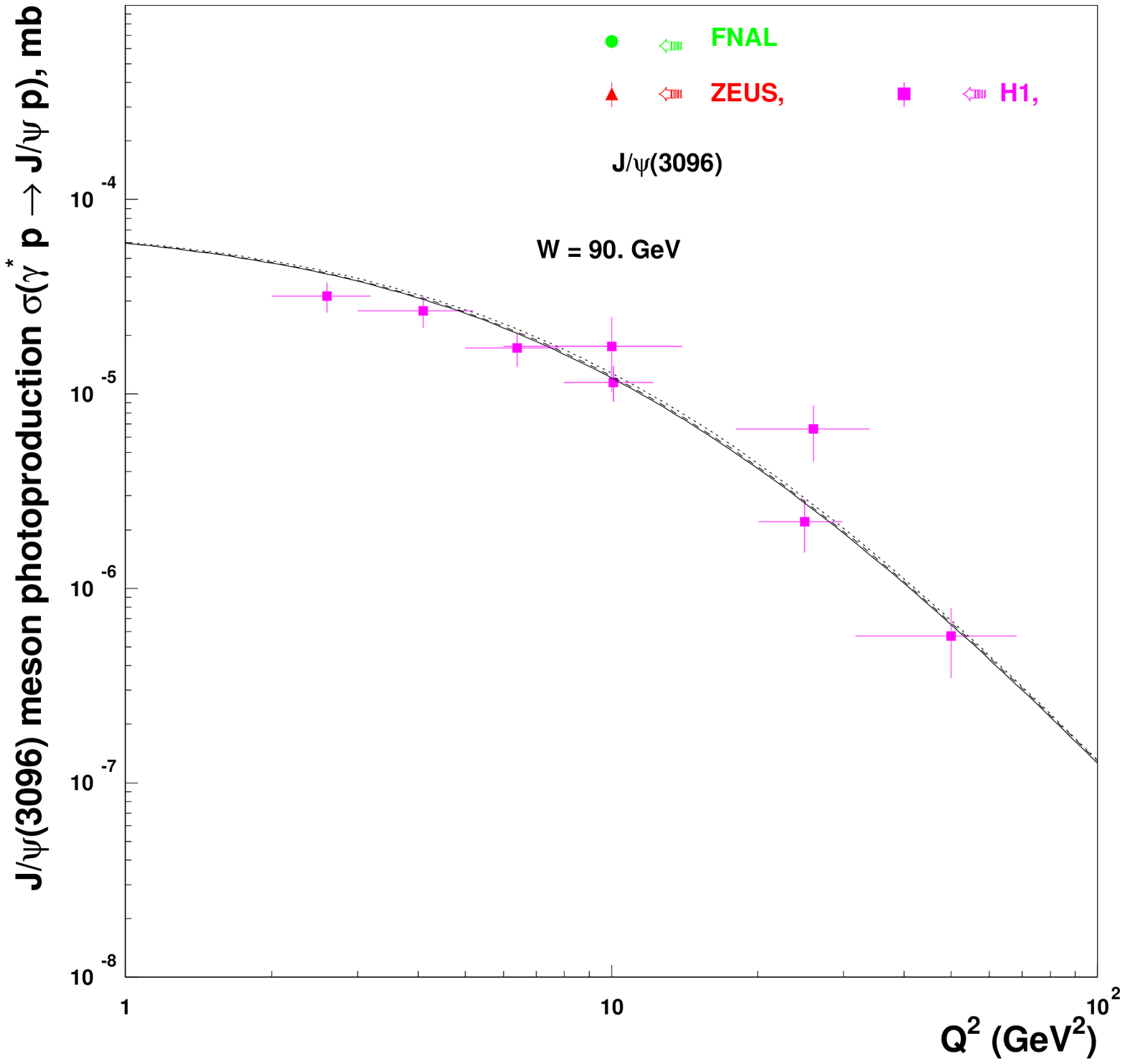}}

\vspace*{-1.3cm}
\parbox[t]{5.cm}{\caption{Elastic cross section of exclusive $J/\psi$
virtual photoproduction as a function of $W$
for various $Q^2$. \label{fig:jpsi}}}
\hfill~\parbox[t]{5.cm}{\caption{Elastic cross section of
exclusive $J/\psi$ virtual photoproduction as a
function of $Q^2$ for $W=90\; GeV$. \label{fig:jpsiq90}}}
\end{figure}

We can now check the predictions of the model. The
$\chi^2/\# point=0.89$ for $J/\psi$ meson exclusive production follows without any
fitting. Both $W$ and $Q^2$ dependences are reproduced very well. Notice that,
so far, the three {\it choices I, II, III} all give equally acceptable reproduction
of the data (see Figs. \ref{fig:jpsi}, \ref{fig:jpsiq90} ).

\begin{figure}[h]
\vspace*{-0.5cm}
\parbox[c]{4.cm}{\epsfxsize=40mm
\epsffile{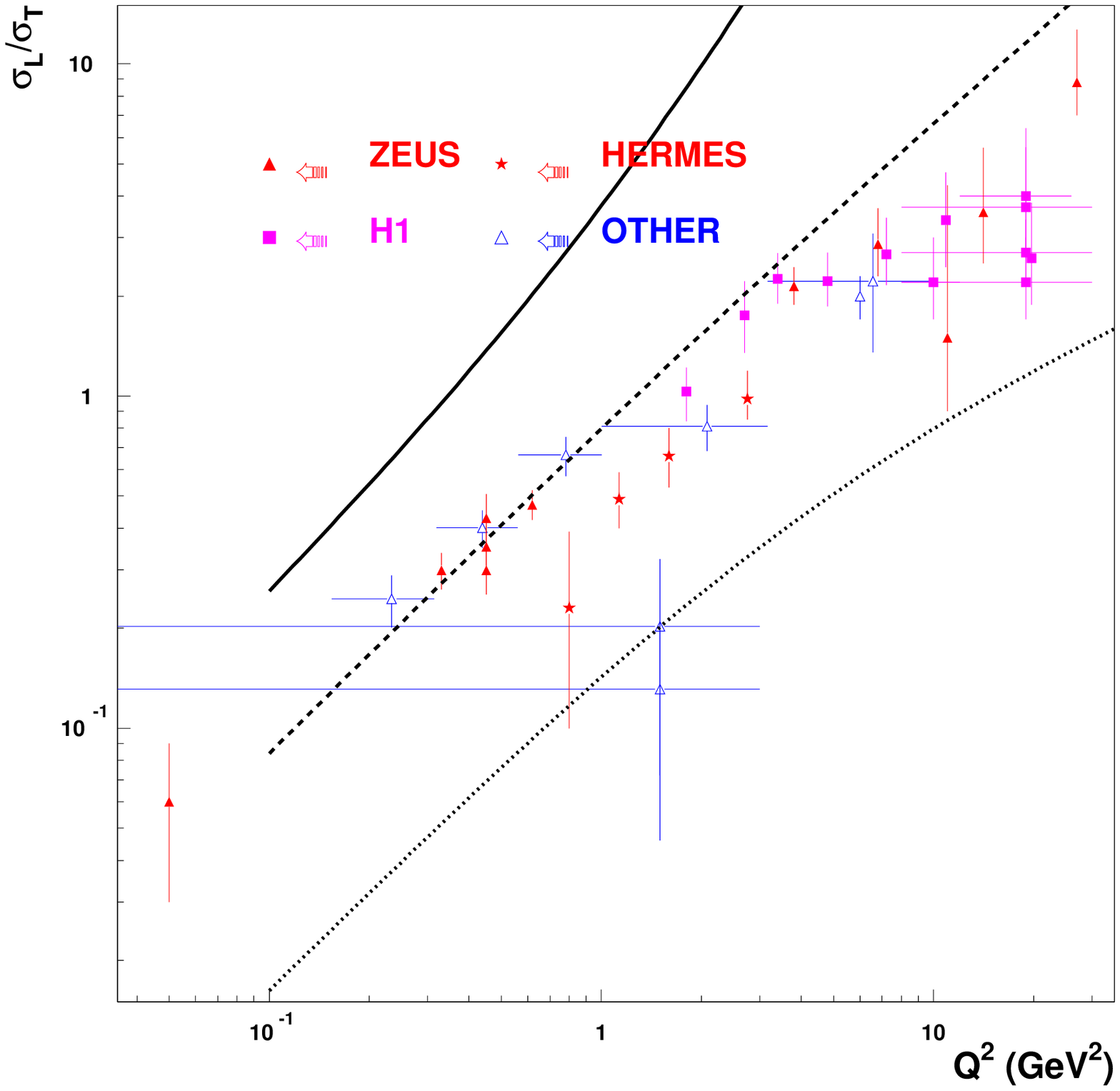}} \hfill~\parbox[l]{6.cm}{\epsfxsize=40mm
\epsffile{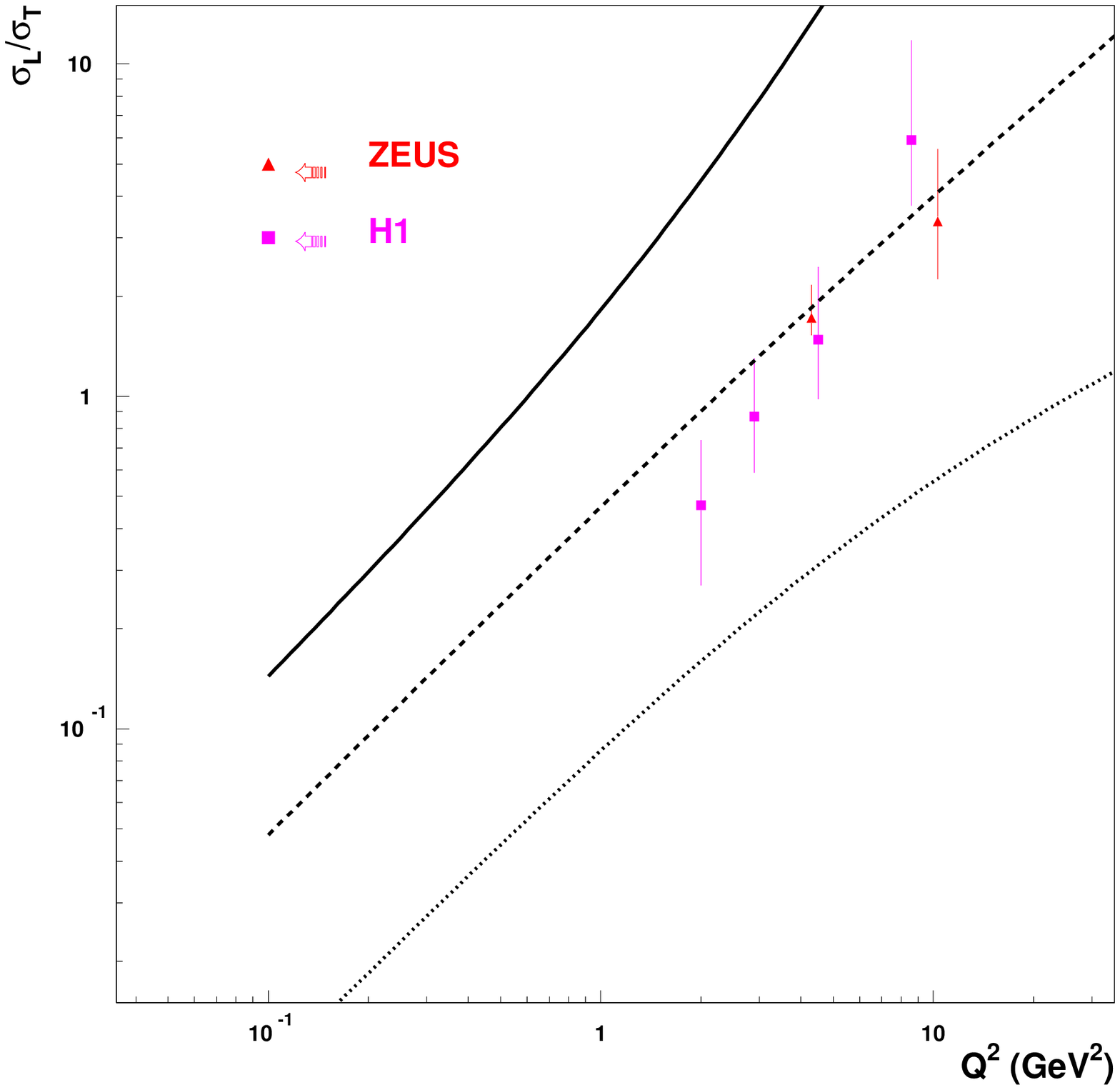}}

\vspace*{-1.cm}
\parbox[t]{5.9cm}{\caption{Ratio of $\sigma_L/\sigma_T$ for exclusive
$\rho_0$ large $Q^2$ photoproduction.
\label{fig:rholt}}}
\hfill~\parbox[t]{5.9cm}{\caption{Ratio of $\sigma_L/\sigma_T$ for
exclusive $\varphi$ large $Q^2$ photoproduction.
\label{fig:philt}}}

\vspace*{-0.3cm}

\parbox[c]{4.cm}{\epsfxsize=40mm
\epsffile{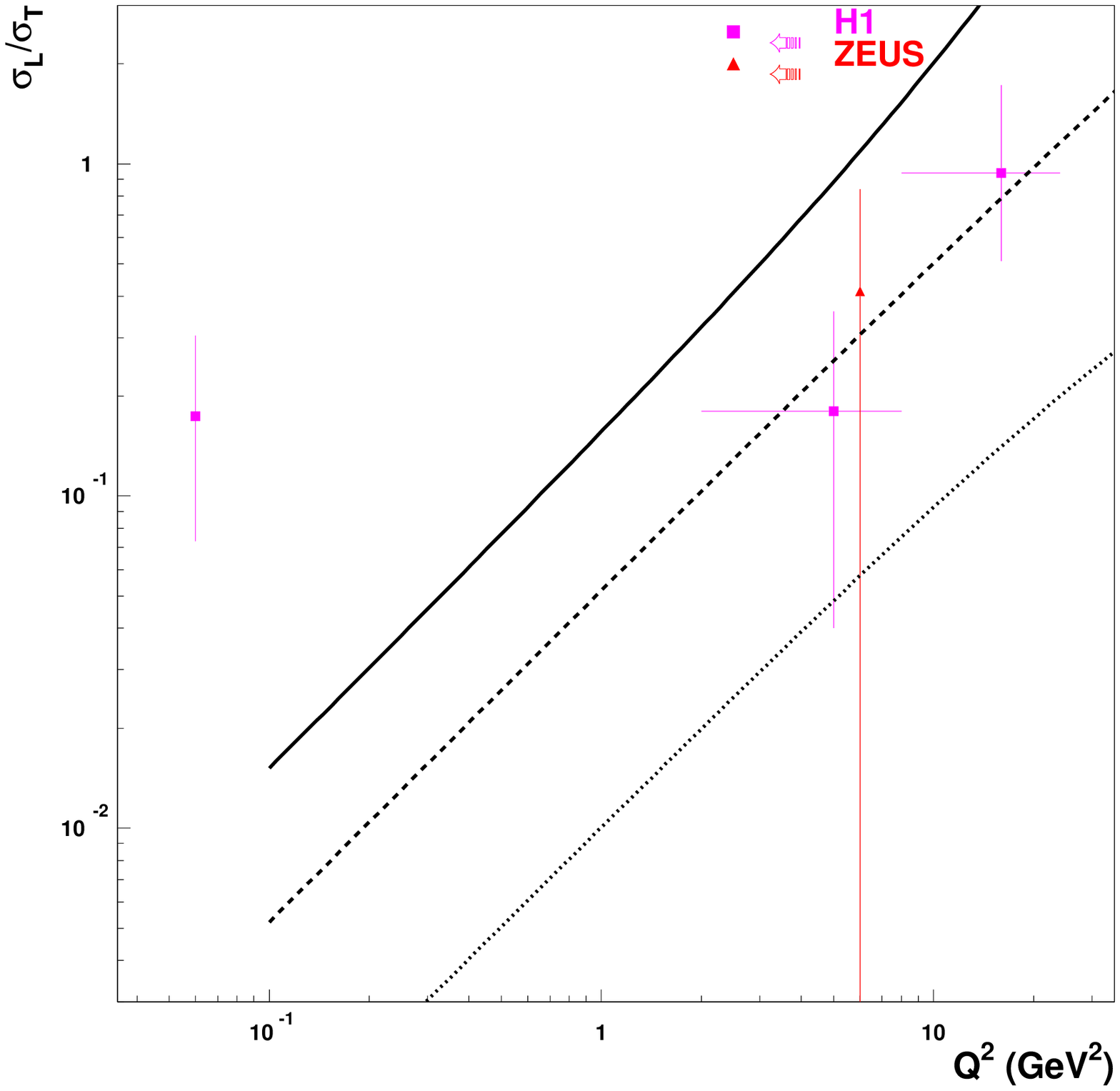}}

\vspace*{-1.cm}
\parbox[t]{5.9cm}{\caption{Ratio of $\sigma_L/\sigma_T$ for exclusive
$J/\psi$ large $Q^2$ photoproduction.
\label{fig:jpsilt}}}
\end{figure}

We now plot the various ratios $\sigma_L/\sigma_T$ (these data were not fitted)
corresponding to
Eqs
(\ref{eq:couplingsChoice}),(\ref{eq:couplings1}),(\ref{eq:couplings2})
(shown with the solid ({\it choice I}), dashed ({\it choice II})
and dotted ({\it choice III}) lines) in Fig. \ref{fig:rholt},
\ref{fig:philt}, \ref{fig:jpsilt}.  The result shows, indeed, a
rapid increase of $\sigma_L/\sigma_T$ with increasing $Q^2$,
however one can see that our intermediate {\it choice II} is preferable to either {\it I} or {\it III} on this basis.

Let us examine the obtained dependences. We find that the data
prefer 
\be
R(Q^2\rightarrow \infty) \sim
\Big(\frac{Q^2}{M_V^2}\Big)^{n_1}\; , 
\ee where $n_1\simeq 2,\;
1,\; 0.3$ in {\it choice I, II and III}. Our, probably
oversimplified, estimates and the data show $0.3<n_1<1$, see
Fig. \ref{fig:rholt}, thus $\sigma\sim 1/Q^N$ where 
$N\in (4,4.4)$ as $N=6-2n_{1}$
for the {\it choice II} and  $N=5-2n_{1}$
for the {\it choice III}. However it is evident that 
new more precise data on $R$ are needed.

\section{Conclusion}

We have shown that the Soft Dipole Pomeron
model recently developed \cite{owrmodel}, \cite{ref:OURPAPER} for vector meson
photoproduction allows us to describe well the new ZEUS data
\cite{NEWZEUS} on the differential and integrated cross-sections
for $\gamma p\to J/\psi p$. Again, all available data on
photoproduction of other vector mesons at $Q^{2}=0$ as well as
$Q^{2}\neq 0$ are well reproduced.

The nonlinear Pomeron trajectory
$\alpha_{P}(t)=1+\gamma\
(\sqrt{4m_{\pi}^{2}}-\sqrt{4m_{\pi}^{2}-t})$ turns out to be more
suitable for the nonlinearity of the diffractive cone
shown by the new ZEUS data. 

We would like to emphasize the following important points
\begin{enumerate}
\item The Pomeron used in the model is a double pole in the complex 
$j$-plane with intercept which is equal to one

\item The new ZEUS data \cite{NEWZEUS} (in contrast to the old ones)
quite definitely point towards the nonlinearity of the Pomeron 
slope and trajectory.

\item Phenomenologically we find that in the region of available
$Q^2$ the ratio $ \sigma_L/\sigma_T  \sim
({Q^2}/{M_V^2})^{n_1}\; , $ where $0.3<n_1< 1\;$. The
definite conclusion can be derived only with new precise
data on the ratio $\sigma_L/\sigma_T$, especially for high
$Q^{2}$.
\end{enumerate}
\subsection*{Acknowledgement}
We would like to thank Michele Arneodo, Alexei Kaidalov, Alexander Borissov,
Jean-Rene Cudell and
Alessia Bruni for various and fruitful
discussions. One of us (E.M.) would like to thank the 
Department of Theoretical Physics of the University of Torino
for its hospitality and financial support during his visit
to Turin.


\end{document}